\documentclass[9pt,twocolumn,twoside]{revtex4-1}

\usepackage{graphicx}
\usepackage{amsmath}
\usepackage{textcomp}
\usepackage{mathrsfs}
\usepackage{amsfonts}

\begin{document}

\newcommand{\vett}[1]{\mathbf{#1}}
\newcommand{\uvett}[1]{\hat{\vett{#1}}}
\newcommand{\beq}{\begin{equation}}
\newcommand{\eeq}{\end{equation}}
\newcommand{\bseq}{\begin{subequations}}
\newcommand{\eseq}{\end{subequations}}
\newcommand{\barr}{\begin{eqnarray}}
\newcommand{\earr}{\end{eqnarray}}
\newcommand{\GH}{Goos-H$\ddot{\mathrm{a}}$nchen }
\newcommand{\IF}{Imbert-Fedorov }
\newcommand{\bra}[1]{\langle #1|}
\newcommand{\ket}[1]{| #1\rangle}
\newcommand{\expectation}[3]{\langle #1|#2|#3\rangle}
\newcommand{\braket}[2]{\langle #1|#2\rangle}

\title{Goos-H\"anchen and Imbert-Fedorov Shifts for Airy Beams}

\author{Marco Ornigotti}

\affiliation{Institut f\"ur Physik, Universit\"at Rostock, Albert-Einstein-Stra\ss e 23, 18059 Rostock, Germany}

\email{marco.ornigotti@uni-rostock.de}

\date{Compiled \today}

\begin{abstract}
In this work, I present a full analytical theory for the Goos-H\"anchen and Imbert-Fedorov shifts experienced by an Airy beam impinging on a dielectric surface. In particular, I will show how the decay parameter $\alpha$ associated with finite energy Airy beams is responsible for the occurence of giant angular shifts. A comparisons with the case of Gaussian beams is also discussed.
\end{abstract}

\maketitle

Accelerating beams, i.e., beams of light that propagate along curved trajectories in free space without any external potential present, attracted a lot of interest in the last years. The forefather of this new class of optical fields is undoubtledy the Airy beam, first introduced in quantum mechanics by Berry and Balazs in 1979 \cite{airy0} and then brought to optics by  by Siviloglou and co-workers in 2007 \cite{airy1,airy2}. This particular solution of the paraxial equation, in fact, propagates along a parabolic trajectory in free space, due to the fact that it experiences a nonzero tangential acceleration in its own comoving reference frame. Driven by the intriguing properties possessed by Airy beams, in the last decade many different classes and types of accelerating beams were proposed, each with unique features, such as the ability of propagating along curved \cite{airy3,airy4}, spiraling \cite{prlNoi} or arbitrary \cite{airy5, airy5bis} trajectories in space, route particles along non-conventional paths \cite{airy6}, or generate curved plasma channels \cite{airyPlasma}, to name a few. Airy beams, in particular, were extensively studied \cite{revAiry}, including a detailed analysis of their reflection from dielectric surfaces close to total internal reflection and Brewster incidence \cite{reflAiry1}. The occurrence of selective reflection was also reported \cite{reflAiry3}. 

Reflection of light upon an interface is a daily witnessed phenomenon, which can be easily understood and explained using Snell's law \cite{ref1}. This law, however, holds exactly only when the light impinging upon a reflecting surface can be described within the geometrical optics approximation. When the wave properties of the impinging field cannot be neglected, like in the case of an optical beam for example, nonspecular reflection phenomena typically occur.  Among these effects, the most famous are certaintly the so-called \GH \cite{ref2,ref3,ref4} and \IF \cite{ref5,ref6,ref7,ref8,ref9,ref10,ref11,ref12,ref13,ref14,ref15} shifts, the former occurring in the plane of incidence and the latter in the plane perpendicular to it. These phenomena have been extensively studied in the past decades for a broad class of beam configurations \cite{ref16,ref17,ref18,ref19,ref20,ref24} and interfaces \cite{ref21,ref22,ref23,ref24bis,ref24ter,ref24quater,ref24quint}. Moroever, the analogy between beam shifts and the quantum mechanical weak measurements has also been proposed \cite{ref26,ref27,ref28,ref29,ref30} and used as a way to get strongly amplified beam shifts \cite{ref31,ref32} . A comprehensive review on the beam shift phenomena can be found in Ref. \cite{ref25}. For the case of Airy beams, intermittend \GH shift from nonlinear surfaces has been predicted and numerically verified in Ref. \cite{reflAiry2}.  However, a comprehensive theory of \GH and \IF shifts for Airy beams has not been developed yet.
\begin{figure}[!t]
\begin{center}
\includegraphics[width=0.5\textwidth]{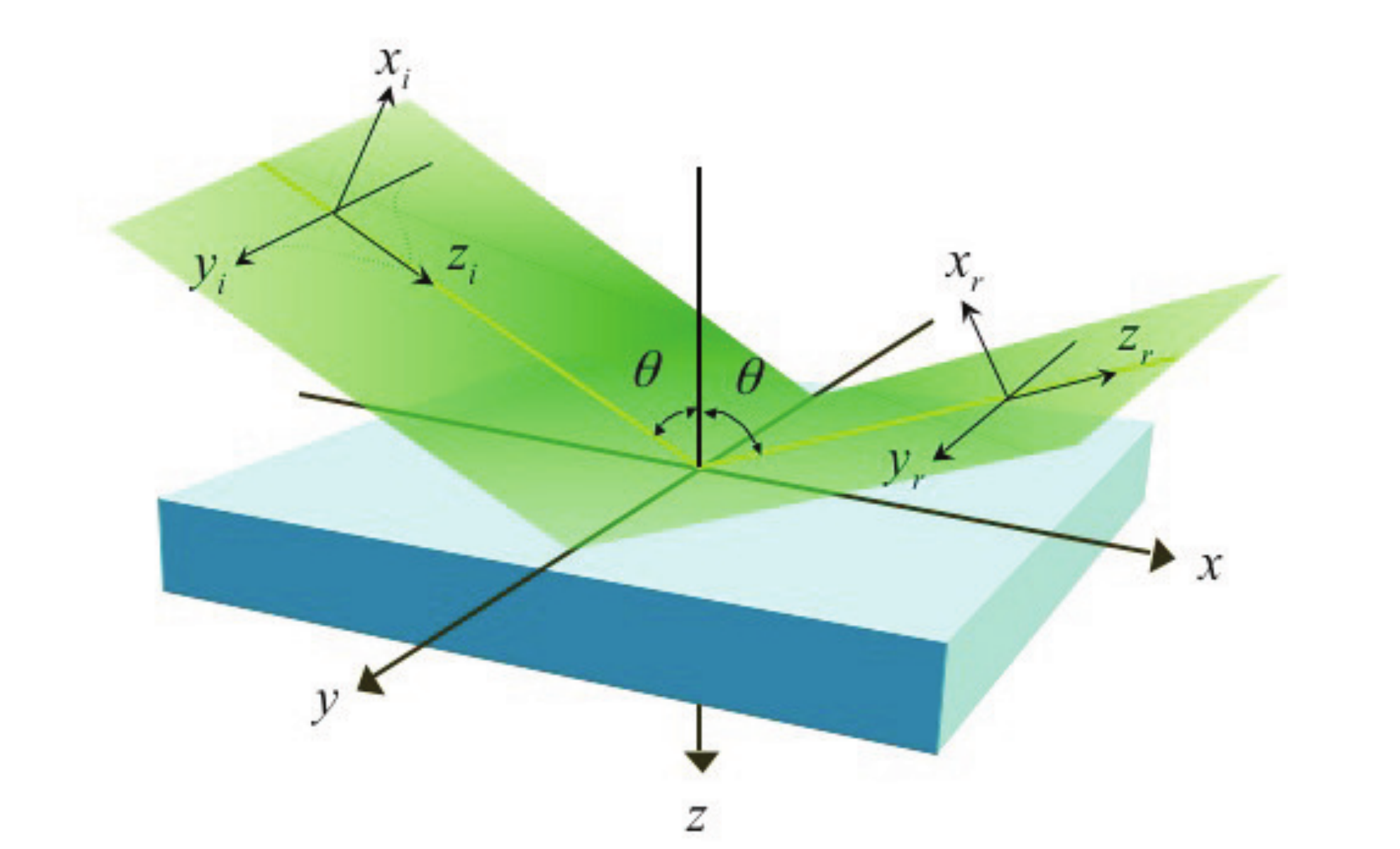}
\caption{(Color online) Schematic representation of the geometry of the problem. The laboratory frame $\{\uvett{x},\uvett{y},\uvett{z}\}$ is attached to the reflecting surface (blue box) , with the $z$-axis pointing downwards. The incident ($\{\uvett{x}_i,\uvett{y}_i,\uvett{z}_i\}$) and reflected ($\{\uvett{x}_r,\uvett{y}_r,\uvett{z}_r\}$) reference frames are instead shown by the green planes. $\theta$ is the angle of incidence.}
\label{figure1}
\end{center}
\end{figure}

In this Letter, therefore, I will present a complete theory of \GH (GH) and \IF (IF) shifts for Airy beams impinging upon a dielectric interface. In particular, I will show that for the case of finite energy Airy beams, giant angular GH and IF shifts occur, as a consequence of the finite width of the Airy beam spectrum. 

To begin with, let us consider a monochromatic, paraxial electric field impinging upon a dielectric surface, as depicted in Fig. \ref{figure1}. For later convenience, we can define three reference frames: the laboratory frame $\{\uvett{x},\uvett{y},\uvett{z}\}$ attached to the dielectric interface and oriented with the $\uvett{z}$ axis pointing towards the dielectric, and two auxiliary frames $\{\uvett{x}_k,\uvett{y}_k,\uvett{z}_k\}$ attached to the incident ($k=i$) and the reflected ($k=r$) field, respectively. The impinging electric field can be then described, in the incident frame, by the following Fourier representation:
\beq\label{eq1}
\vett{E}_{inc}(\vett{r})=\sum_{\lambda=1}^2\int d^2K\uvett{e}_{\lambda}(U,V;\theta)A_{\lambda}(U,V)e^{i(UX_i+VY_i+WZ_i)},
\eeq
where $d^2K=dUdV$, $\lambda$ is a polarization index ($\lambda=1$ corresponds to $p$-polarization and $\lambda=2$ corresponds to $s$-polarization). In the expression above, $U$, $V$ and $W=\sqrt{1-U^2+V^2}\simeq 1-(U^2+V^2)/2$ are the dimensionless components of the $\vett{k}$-vector in the incident frame, which are connected to the $\vett{k}$-vector in the laboratory frame by the relation $\vett{k}=k_0\left(U\uvett{x}_i+V\uvett{y}_i+W\uvett{z}_i\right)=k_x\uvett{x}+k_y\uvett{y}+k_z\uvett{z}$ \cite{ref25}.  Moreover, $A_{\lambda}(U,V)=\alpha_{\lambda}(U,V;\theta)A(U,V)$ is the vector spectral amplitude, with $A(U, V)$ being the angular spectrum of the incident field and  $\alpha_{\lambda}(U,V)$  the polarisation functions given by \cite{andrea1}
\beq\label{eq2}
\alpha_{\lambda}(U,V)=\uvett{f}\cdot\uvett{e}_{\lambda}(\vett{k}),
\eeq
with $\uvett{e}_{\lambda}(\vett{k})$ being the polarisation vectors attached to the single plane wave components of the impinging beam, whose definition is given as $\uvett{e}_1(\vett{k})=\left(\uvett{e}_2\times\vett{k}\right])/|\uvett{e}_2\times\vett{k}|$ and $\uvett{e}_2(\vett{k})=(\uvett{z}\times\vett{k})/|\uvett{z}\times\vett{k}|$. The unit vector $\uvett{f}=a_p\uvett{x}_i+a_se^{i\eta}\uvett{y}_i$ (with $a_p,a_s,\eta\in\mathbb{R}$) describes the polarization state of the beam in the incident frame. Upon reflection, each plane wave component of the Fourier spectrum in \eqref{eq1} gets reflected according to Fresnel's law, i.e., 
\beq\label{eq4}
\uvett{e}_{\lambda}(\vett{k})e^{i\vett{k}\cdot\vett{r}}\hspace{0.2cm}\rightarrow\hspace{0.2cm}r_{\lambda}(\vett{k})\uvett{e}_{\lambda}(\tilde{\mathbf{k}})e^{i\tilde{\mathbf{k}}\cdot\vett{r}},
\eeq
where $r_{\lambda}(\vett{k})$ are the Fresnel reflection coefficients \cite{ref1} and $\tilde{\mathbf{k}}=\vett{k}-2\uvett{z}\left(\uvett{z}\cdot\vett{k}\right)$ is determined by the law of specular reflection \cite{gragg}. The expression for the electric field $\vett{E}_{refl}(\vett{r})$ in the reflected frame can be the obtained from \eqref{eq1} upon substituting $\uvett{e}_{\lambda}(\vett{k})\rightarrow\uvett{e}_{\lambda}(\tilde{\vett{k}})$ and $\vett{k}_i\rightarrow\vett{k}_r=-U\uvett{x}+V\uvett{y}+W\uvett{z}$, thus obtaining
\beq\label{reflField}
\vett{E}_{refl}(\vett{r})=\sum_{\lambda=1}^2\int d^2K\,\mathcal{E}(U,V;\theta)e^{i(-UX_r+VY_r+WZ_r)}.
\eeq
where $\mathcal{E}_{\lambda}(U,V;\theta)=\uvett{e}_{\lambda}(-U,V;\pi-\theta)r_{\lambda}(U,V)\alpha_{\lambda}(U,V;\theta)A(U,V)$ contains informations about both the reflection coefficients and the angular spectrum of the impinging beam. According to the standard theory of beam shifts \cite{ref20, andrea2}, to evaluate the GH and IF shifts one first needs to calculate the centroid of the intensity distribution of the reflected field, i.e., $\langle\vett{X}\rangle=\langle X_r\rangle\uvett{x}_r+\langle Y_r\rangle\uvett{y}_r$, where
\beq\label{centroid}
\langle X_r\rangle=\frac{\int\,d^2R\,X_r\,I(R,z)}{\int\,d^2R\,I(R,z)}\nonumber\\
\eeq
being $d^2R=dX_rdY_r$, $I(R,z)=\vett{E}_{refl}(R,z)\cdot\vett{E}_{refl}(R,z)$ is the intensity distribution of the reflected field in the reflected frame.By substituting \eqref{reflField} in the above expression, after some straightforward calculations we obtain the following expresison for the centroid
\barr\label{centroidK}
\langle X_r\rangle&=&\frac{1}{\mathcal{N}}\operatorname{Im}\Big\{\int\,d^2K\,\mathcal{E}^*(U,V;\theta)\frac{\partial}{\partial U}\mathcal{E}(U,V;\theta)\nonumber\\
&-&Z\int\,d^2K\,\frac{U}{W}\left|\mathcal{E}(U,V;\theta)\right|^2\Big\},
\earr
where $\mathcal{N}=\int\,d^2K\,\left|\mathcal{E}(U,V;\theta)\right|^2$, and $d^2K=dUdV$. An analogue expression (with the substitution $\partial/\partial U\rightarrow\partial/\partial V$ and $U/W\rightarrow V/W$) is valid for $Y_r$. Then, the spatial ($\Delta$) and angular ($\Theta$) GH and IF shifts have the following form:
\bseq
\begin{align}
\Delta_{GH}&=\langle X_r\rangle|_{z=0}, \hspace{1cm} \Theta_{GH}&=\frac{\partial\langle X_r\rangle}{\partial z},\\
\Delta_{IF}&=\langle Y_r\rangle|_{z=0}, \hspace{1cm} \Theta_{IF}&=\frac{\partial\langle Y_r\rangle}{\partial z}.
\end{align}
\eseq
In order to express the GH and IF shifts for Airy beams in a compact form, it is useful to first recall the expressions for the GH and IF shifts for a Gaussian beam impinging upon a dielectric surface. These can be obtained by setting $A(U,V)$ in \eqref{eq1} to be a Gaussian function and performing the integrations in \eqref{centroid} by expanding the integrand functions in the numerator and denominator up to the first order in $U$ and $V$ (by virtue of the paraxial approximation) for the spatial shifts and the numerator up to the second order for the angular shifts. After some manipulations, we get the following well known expressions \cite{ref20}:
\bseq\label{gaussGHIF}
\begin{align}
\Delta_{GH}^{(g)}&=\frac{1}{k_0}\left(w_p\frac{\partial\phi_p}{\partial\theta}+w_s\frac{\partial\phi_s}{\partial\theta}\right),\\
\Theta_{GH}^{(g)}&=\frac{1}{k_0z_R}\left(w_p\frac{\partial\ln R_p}{\partial\theta}+w_s\frac{\partial\ln R_s}{\partial\theta}\right),\\
\Delta_{IF}^{(g)}&=-\frac{1}{k_0}\Big\{\cot\theta\Big[\frac{w_pa_s^2+w_sa_p^2}{a_pa_s}\cos\eta\nonumber\\
&+ 2\sqrt{w_pw_s}\sin(\eta+\phi_s-\phi_p)\Big]\Big\},\\
\Theta_{IF}^{(g)}&=\frac{1}{k_0z_R}\left[\frac{w_pa_s^2-w_sa_p^2}{a_pa_s}\cos\eta\cot\theta\right],
\end{align}
\eseq
where the superscript $^{(g)}$ stands for Gaussian beam, $k_0=2\pi/\lambda$ and $z_R$ is the Rayleigh range of the beam . In the above expresions, $R_{\lambda}$ and $\phi_{\lambda}$ are the modulus and phase of the reflection coefficients, i.e., $r_{\lambda}=R_{\lambda}e^{i\phi_{\lambda}}$ \cite{andrea1} and $w_{\lambda}=a_{\lambda}^2R_{\lambda}^2/(a_p^2R_p^2+a_s^2R_s^2)$ is the fractional energy contained in each polarisation.

For the case of Airy beams, instead, the angular spectrum appearing in \eqref{eq1} has the following form \cite{airy0}:
\beq\label{eq3a}
A(U,V)=e^{\frac{i}{3}(U^3+V^3)},
\eeq
which corresponds, in the plane $z=0$, to the 2D pure (infinite energy) Airy beam $E(x,y,z=0)=\text{Ai}(x)\text{Ai}(y)$. To calculate the beam shifts for such field distribution, therefore, one should use \eqref{eq3a} to calculate the integrals appearing in \eqref{centroid}. In doing so, however, one needs to deal with integrals of the type
\beq
\mathcal{I}(n,m)\equiv\int\,dU\,dV\,U^n\,V^m\left|A(U,V)\right|^2,
\eeq
where $n,m\in\mathbb{N}$. With the choice of angular spectrum as the one given by \eqref{eq3a}, it is not difficult to show that $\mathcal{I}(n,m)=\infty$. This result prevents one to calculate the integrals appearing in \eqref{centroidK}, and, moreover, their ratio. To overcome this problem, we can introduce a regularisation parameter in the integrals $\mathcal{I}(n,m)$, in such a way that they give finite results, thus allowing a correct evaluation of \eqref{centroid}. Then, by taking the limit of the regularised result when the regularisation parameter goes to zero, we can obtain the correct result for pure Airy beams. Conceptually, this regularisation technique is the same one employed for the calculation of \GH and \IF shifts for Bessel beams \cite{ref24}.

A convenient regularisation condition for Airy beams, is the one given in Ref. \cite{airy1}, which corresponds to the so-called finite energy Airy beams. This regularisation consists in adding a proper exponential factor to the expression of the Airy beam, namely $E(x,y,z=0)=\text{Ai}(x)\text{Ai}(y)\exp{[\alpha(x+y)]}$. The regularisation parameter $\alpha>0$ ensures a containment of the tails of the Airy function, thus allowing such beams to carry finite energy and be then physically realisable \cite{airy1}. The angular spectrum of such a regularised beam can be calculated using the integral representation of the Airy function \cite{airyBook}
\beq
\text{Ai}(x)=\frac{1}{2\pi}\int\,dz\,e^{i\left(\frac{z^3}{3}+xz\right)},
\eeq
and the identity \cite{airyBook}
\beq
\int_{-\infty}^{\infty}\,dx\,e^{xy}\text{Ai}(x)=e^{\frac{y^3}{3}}.
\eeq
This gives the following result:
\beq\label{eq3}
A(U,V)=e^{-\alpha^2(U^2+V^2)+\frac{\alpha^3}{3}}e^{\frac{i}{3}\left[(U^3+V^3)-3\alpha^2(U+V)\right]}.
\eeq
In the limit $\alpha=0$, the above angular spectrum reduces to \eqref{eq3a}. Notice, moreover, that $|A(U,V)|^2=\exp{[-2\alpha^2(U^2+V^2)]}\exp{[2\alpha^3/3]}$. Therefore, the integrals in \eqref{centroid} become Gaussian integrals. This suggests that the GH and IF shifts for regularised Airy beam will have a form similar to the one for Gaussian beams.

The beam shifts for an Airy beam can be then calculated using the regularised angular spectrum given by \eqref{eq3} (which, then, corresponds to finite energy Airy beams) and then take the limit of the results obtained in this way when $\alpha\rightarrow 0$ to obtain the correspondent result for pure (infinite energy) Airy beams. If we now calculate the integrals in \eqref{centroid}  for the case of regularised Airy beams, we obtain the following result:
\bseq\label{AiryShift}
\begin{align}
\Delta_{GH}&=\Delta_{GH}^{(g)}-\Gamma,\hspace{0.2cm}\Delta_{IF}=\Delta_{IF}^{(g)}+\Gamma,\label{AiryShiftc}\\
\Theta_{GH}&=\frac{1}{2\alpha^2}\Theta_{GH}^{(g)},\hspace{0.2cm}\Theta_{IF}=\frac{1}{2\alpha^2}\Theta_{IF}^{(g)},\label{AiryShiftd}
\end{align}
\eseq
where $\Gamma=\alpha^2/k_0$. This is the main result of this Letter. Before going any further, it is worth commenting on the results pre- sented above. To understand them, one needs first to recall, that when calculating the beam shifts, both the numerator and the denominator of  \eqref{centroid} are Taylor expanded in $U$ and $V$ up to the first order. This means that only terms proportional to$\mathcal{I}(0,0)$, $\mathcal{I}(1,0)$, and $\mathcal{I}(0,1)$ will contribute to define the shifts.  For the case of the spectrum in \eqref{eq3a}, moreover, since $|A(U,V)|^2$ is a Gaussian function of $U$ and $V$, $\mathcal{I}(1,0)=\mathcal{I}(0,1)=0$, and therefore only terms proportional to $\mathcal{I}(0,0)$ will contribute in determining the spatial and angular shifts in \eqref{AiryShift}. This justifies the fact that the shifts for regularised Airy beams looks very similar to the ones for Gaussian beams. For the angular shifts, instead, the presence of the factor $U/W=U/\sqrt{1-U^2-V^2}$ in \eqref{centroidK} requires the integrals to be expanded up to second order in $U$ and $V$. Therefore, the result will be proportional to $\mathcal{I}(2,0)/\mathcal{I}(0,0)\propto 1/\alpha^2$.	

To understand the presence of the $\Gamma$-factor in the spatial shifts, instead, one should note that when calculating the derivative in \eqref{centroidK}, the term $\partial A(U,V)/\partial U$ (or its $V$-counterpart) gives a contribution of the form $(-2\alpha^2U+iU^2-i\alpha^2)$. Since only $\mathcal{O}(U^2,V^2)$ terms must be taken into account, and Gaussian integrals with odd powers vanish, only the term proportional to $\alpha^2$ will survive, thus giving the extra $\Gamma$ term in  Eqs. (\ref{AiryShiftc}). Notice, that this term comes from the extra linear phase $\exp{[-i\alpha(U+V)]}$ present in the spectrum because of the regularisation. As $\alpha\rightarrow 0$, therefore, this contribution vanishes, as the angular spectrum for a pure Airy beam does not contain such a linear phase term. For infinite energy Airy beams, therefore, the spatial \GH and \IF shifts resemble the ones for a Gaussian beam. 

For finite energy Airy beams, the decay factor $\alpha$ affects both the spatial and angular shifts. For the spatial shifts, it  accounts for an additive quantity  $\pm\Gamma$. For the angular shifts, on the other hand, the decay factor appears as a multiplicative constant in the angular shifts, thus resulting in an amplification or suppression of the shift (with respect to the case of a Gaussian beam), depending on whether  $\alpha<1$  or $\alpha>1$, respectively. In practical cases, according to Refs. \cite{airy1,airy2}, $\alpha$ is chosen to be a small number, of the order of $10^{-2}$. This choice is motivated by the necessity of having a rapidly enough decaying function, which ensures the Airy beam to carry finite energy. In this case, therefore, $\Gamma\ll1$ and the spatial shifts are the ones of a Gaussian beam. This is in accordance with the numerical results obtained in in Ref. \cite{reflAiry3}, where the predicted spatial GH shift is essentially the one of a Gaussian beam [see for example Fig. 5(c) of  Ref. \cite{reflAiry3}]. For the angular shifts, instead, we have $\Theta_{GH,IF}/\Theta_{GH,IF}^{(g)}=1/2\alpha^2\simeq 10^3$. Finite energy Airy beams thus display giant angular GH and IF shifts. The amplitude of such angular shifts can be then controlled by suitably tuning the decay paramenter $\alpha$. 
\begin{figure}[!t]
\begin{center}
\includegraphics[width=0.5\textwidth]{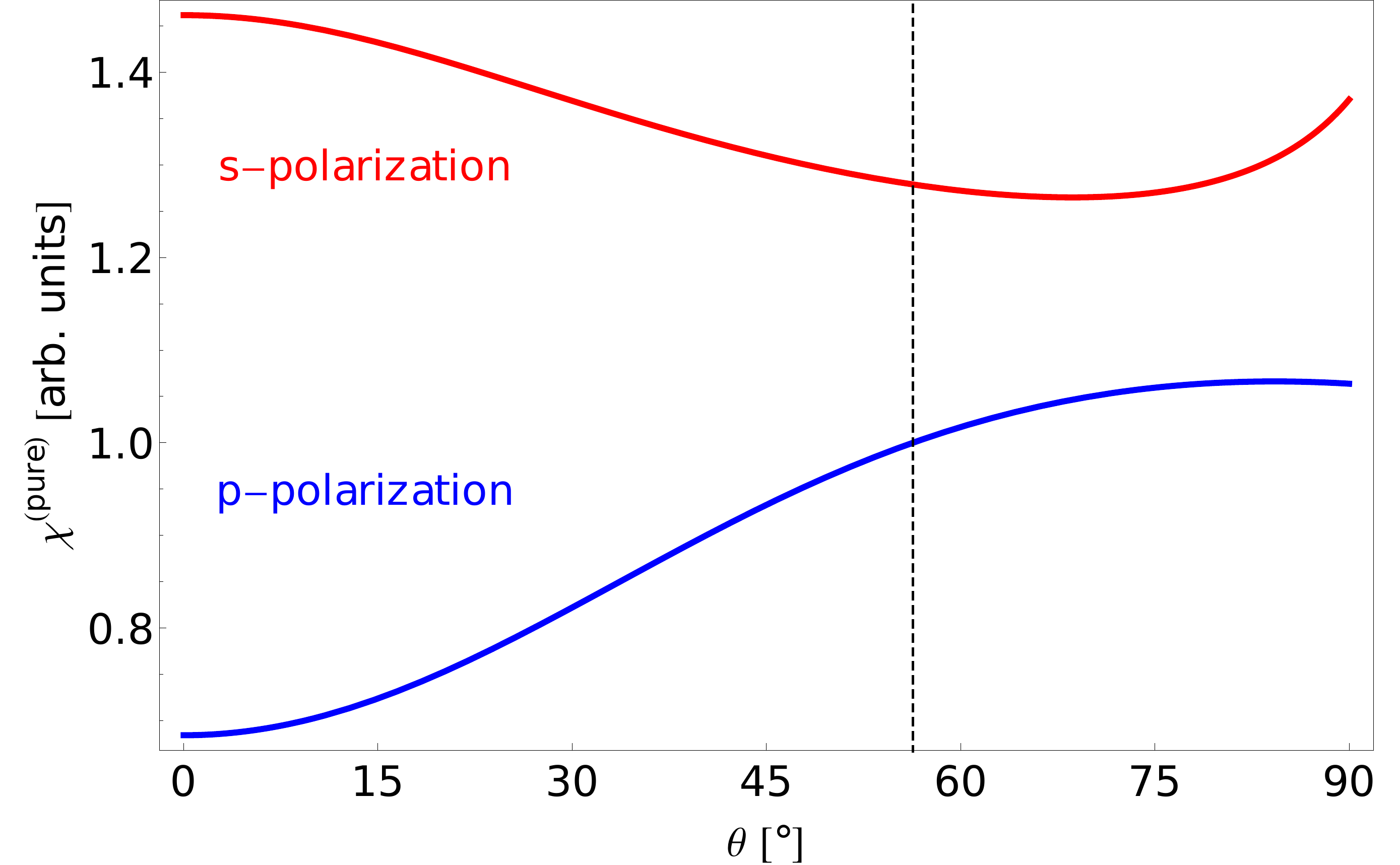}
\caption{Amplification factor $\chi^{(pure)}$ for a pure Airy beam for an air-glass interface ($n_{glass}=1.5$) for $p$- (blue, lower, solid line) and $s$-polarisation (red, upper, solid line). As it can be seen, while $\chi^{(pure)}_p<1$ for $\theta<\theta_B$, $\chi_s^{(pure)}>1$ for every incident angle. The black, dashed vertical line indicates the position of the Brewster angle, which in this case is $\theta_B\simeq 56,3^{\circ}$.}
\label{figure2}
\end{center}
\end{figure}

As a last remark, it is also interesting to explore the limiting case of \eqref{AiryShift} when $\alpha\rightarrow 0$, corresponding to GH and IF shifts for nondiffracting, infinite energy Airy beams. For the case of spatial shifts, this limit is immediate, as $\Gamma\rightarrow 0$ when $\alpha\rightarrow 0$, thus meaning that the spatial shifts for infinite energy Airy beams are the same as for a Gaussian beam. For the angular shifts, on the other hand, the presence of $\alpha$ at the denominator makes the limit of Eqs. (\ref{AiryShiftd}) diverge. This is due to the fact that, while the integral in the numerator of the angular part of the shifts is expanded up to second order in $U$ and $V$, the denominator is only expanded up to first order in $U$ and $V$. A similar problem, for example, occurs as well when dealing with bean shifts close to Brewster incidence \cite{andrea1}. To solve this problem, it is then sufficient to calculate the normalisation factor appearing in \eqref{centroidK} up to second order precision in $U$ and $V$. By doing so, we obtain $\Theta_{GH}=\chi\,\widetilde\Theta_{GH}^{(g)}$, and $\Theta_{IF}=\chi\,\widetilde\Theta_{IF}^{(g)}$,
(where $\widetilde\Theta_{GH,IF}^{(g)}$ are obtained from Eqs. \eqref{gaussGHIF} by substituting $w_{\lambda}$ with $\tilde{w}_{\lambda}=a_{\lambda}^2R_{\lambda}^2/[a_p^2(R_p^2+\varepsilon_p)+a_s^2(R_s^2+\varepsilon_s)]$). The second order corrected amplification factor $\chi$ is given by
\beq
\chi=\frac{a_p^2(R_p^2+\varepsilon_p)+a_s^2(R_s^2+\varepsilon_s)}{a_p^2(2R_p^2\alpha^2+\varepsilon_p)+a_s^2(2R_s^2\alpha^2+\varepsilon_s)},
\eeq 
where $\varepsilon_{\lambda}=(1/2)[(R_p')^2+R_pR_p''-R_s^2+R_pR_p'\cot\theta-(R_p^2-R_s^2)\csc^2\theta]$ is the second-order correction to the total energy carried by the beam \cite{andrea1}, and the prime indicates derivation with respect to $\theta$.

The expression above for the amplification factor represents the correct form to use in the vicinity of Brewster incidence, and allows us to correctly take the limit $\alpha\rightarrow 0$. For linear polarisation, in particular, we get $\chi_{\lambda}^{(pure)}=1+R_{\lambda}^2/\varepsilon_{\lambda}$, ($\lambda=\{p,s\}$). In particular, if we consider $p$-polarisation, at the Brewster angle we have $R_p=0$ and we therefore obtain $\chi_p^{(pure)}=1$. As can be seen from Fig. \ref{figure2}, moreover, the amplification factor is greater than one for every incident angle only for $s$-polarisation (red, upper, solid line), while for $p$-polarisation it is greater than one only for $\theta>\theta_B$.

In conclusion, I have calculated GH and IF shifts for both finite and infinite energy Airy beams. For the former, the decay factor $\alpha$ is responsible for the occurrence of a giant angular GH and IF shift, which, for realistic values of $\alpha$ are predicted to be approximately a thousand times bigger than the correspondent shifts for a Gaussian beam. Moreover, finite energy Airy beams possess nonzero $\Delta_{GH}$ and $\Delta_{IF}$ shifts even when no total internal reflection takes place. This is in contrast with usual results for Gaussian beams \cite{ref24ter, ref24quater,ref24quint}, where $\Delta_{GH}\neq 0$ only close to total internal reflection. This residual spatial shift, however, is significantly smaller than the wavelength, as  $\Gamma\propto\alpha^2\lambda\ll\lambda$ for realistic values of $\alpha$. For the case of infinite energy carrying Airy beams, instead, the angular shifts are enhanced by a quantity $\chi$, which is independent on the form of the beam and selectively amplifies $s$-polarisation at all incident angles. The spatial shifts $\Delta_{GH,IF}$, on the other hand, are unaffected by the particular structure of Airy beams and correspond to the ones for Gaussian beams.

The author would like to thank A. Szameit and L. Ornigotti for stimulating discussions and careful proofreading of the manuscript.

\end{document}